\documentclass[aps, twocolumn,  superscriptaddress]{revtex4-2}

\usepackage{multirow} 
\usepackage{booktabs}
\setcitestyle{super}
\usepackage{graphicx}
\usepackage{amssymb}

\usepackage{amsmath}
\usepackage{xspace}
\usepackage{longtable}
\usepackage{soul}
\usepackage{xcolor}
\usepackage{epstopdf}
\usepackage{array}
\usepackage{tabularx}
\usepackage{physics}
\usepackage{float}
\usepackage{comment}
\usepackage{dsfont}
\usepackage[utf8]{inputenc}
\usepackage{siunitx}

\usepackage[colorlinks = true,linkcolor = blue,urlcolor  = blue,citecolor = blue,anchorcolor = blue]{hyperref}

\usepackage{etoolbox}
\patchcmd{\section}
  {\centering}
  {\raggedright}
  {}
  {}
\patchcmd{\subsection}
  {\centering}
  {\raggedright}
  {}
  {}
\usepackage[compact]{titlesec}         

\AtBeginDocument{
  \setlength\abovedisplayskip{0pt}
  \setlength\belowdisplayskip{0pt}}

\titlespacing*{\subsection}
  {0pt}{0\baselineskip}{0\baselineskip}

\newcommand{\HoW}{HoW$_{10}$\xspace}
\newcommand{\Efield}{$E$-field\xspace}
\newcommand{\df}{$\delta f$\xspace}
\newcommand{\HoWF}{$[$Ho(W$_5$O$_{18}$)$_2]^{9-}$\xspace}

\begin{document}

\title{A call for frugal modelling: two case studies involving molecular spin dynamics}
\author{Gerliz M. Gutiérrez-Finol}
\affiliation{Instituto de Ciencia Molecular (ICMol), Universitat de Val\`encia, Paterna, Spain}

\author{Aman Ullah }
\affiliation{Instituto de Ciencia Molecular (ICMol), Universitat de Val\`encia, Paterna, Spain}

\author{María González-Béjar}
\affiliation{Instituto de Ciencia Molecular (ICMol), Universitat de Val\`encia, Paterna, Spain}

\author{Alejandro Gaita-Ariño}
\email{alejandro.gaita@uv.es}
\affiliation{Instituto de Ciencia Molecular (ICMol), Universitat de Val\`encia, Paterna, Spain}

\date{\today}

\begin{abstract}
As scientists living through a climate emergency, we have a responsibility to lead by example, or to at least be consistent with our understanding of the problem. This common goal of reducing the carbon footprint of our work can be approached through a variety of strategies. For theoreticians, this includes not only optimizing algorithms and improving computational efficiency but also adopting a frugal approach to modeling. Here we present and critically illustrate this principle. First, we compare two models of very different level of sophistication which nevertheless yield the same qualitative agreement with an experiment involving electric manipulation of molecular spin qubits while presenting a difference in cost of $>4$ orders of magnitude. As a second stage, an already minimalistic model of the potential use of single-ion magnets to implement a network of probabilistic p-bits, programmed in two different programming languages, is shown to present a difference in cost of a factor of $\simeq 50$. In both examples, the computationally expensive version of the model was the one that was published. As a community, we still have a lot of room for improvement in this direction.
\end{abstract}

\maketitle


\section*{Introduction: doing science during a climate crisis}

We live in a society of growth. As a positive result, we have obtained accelerated advances in science and technology that have enabled improved welfare and human development. A negative side result one cannot overlook, however, is anthropogenic climate change, a challenge that is undoubtly deserving of our focused attention. From the technological point of view, innovation and efficiency are being studied as tools to decrease CO$_2$ emission intensity, and indeed one can expect that this will play an important role in our response to the climate crisis.\cite{Wenlong2023, Long2018, Kwang2017, Dauda2021} There are also voices that instead call either for limits to growth or even for degrowth, and this point of view will also likely be relevant going forward.\cite{limits, Hickel2022,Slamerak2024} Here we propose to reconcile both views and put them in the context of chemistry research, and in particular in the modelling of chemical systems of interest in Physics and Materials Science.

The role of science in this crisis has been twofold: quantifying and explaining the processes, and also pointing towards our possible ways out. Abundant datatasets have been employed to estimate key indicators related to forcing of the climate system, including emissions of greenhouse gases and short-lived climate forcers, greenhouse gas concentrations, radiative forcing, surface temperature changes, the Earth’s energy imbalance, warming attributed to human activities, the remaining carbon budget, and estimates of global temperature extremes. In a never-ending string of records, Copernicus Climate Change Service reported that June 2024 was the thirteenth month in a row that was the warmest in the ERA5 data record for the respective month of the year.
Indeed, over the 2013–2022 period, human-induced warming has been increasing at an unprecedented rate of over 0.2 $^\circ$C per decade \cite{essd2023}. 

For a more solid perspective, let us explicitly ground our arguments on the Intergovernmental Panel on Climate Change (IPCC). Recent IPCC reports point out how digital technologies, analytics and connectivity consume large amounts of energy, implying higher direct energy demand and related carbon emissions. Thus, the demand of computing services increased by 550\% between 2010 and was by 2022 estimated at 1\% of global electricity consumption. Climate concerns have prompted the development and implementation of green computing policies, which refers to the environmentally responsible use of computers and related technology, focusing on minimizing the environmental impact of information technology (IT) operations by promoting energy efficiency, reducing waste, and encouraging sustainable practices throughout the lifecycle of computer systems. This includes many strategies that take into account designing energy-efficient hardware, utilizing cloud computing, adopting virtualization, recycling electronic waste, and implementing eco-friendly policies in IT environments. All of this aiming to reduce the carbon footprint of computing activities and promote sustainability in the tech industry. Despite the green computing policies implemented by chip manufacturers and computer companies the increase in demand is not quite compensated by efficiency improvements, resulting in an energy demand rising by 6\% between 2000 and 2018.\cite{IPCC2022tecsummary}
This is an example of the uneven policy coverage that exists (high confidence) across sectors: policies implemented by the end of 2020 were projected to result in higher global greenhouse gases emissions in 2030 than emissions implied by the Nationally Determined Contributions (high confidence). Without a strengthening of policies, global warming of \SI{3.2}{\celsius} [2.2 to \SI{3.5}{\celsius}] by 2100 was projected (medium confidence).\cite{IPCC2023syntreport} 
As the potential of demand-side mitigations is considered, a 73\% reduction in electricity use (before additional electrification) is considered by 2050.\cite{IPCC2023syntreport} To the best of our knowledge, IPCC reports have not yet estimated the projected growth of the electrical demand in computing due to the so-called "generative artificial intelligence" although informal estimates, extrapolating from recent trends, anticipate a growth that is incompatible with any reasonable climate goal.

Of course, the most relevant decisions towards mitigation and adaptation are in the hands of policymakers and mostly outside our hands for those of us working in the academia, and yet an obvious question is: do we need to do any changes in our academic policies and in our day-to-day work?
The first aspect in which the academia has answered ``yes'' concerns air travel, and there are some ongoing efforts to reduce our collective carbon footprint in that direction. \cite{Grlinger2023}
Indeed, an extensive internal study at the Institute for Chemical Research of Catalonia resulted in an estimated 48 tons of equivalent CO$_2$ footprint per centre user in 2022, where business-related travel accounted for over 80\% of the centre's emissions. This is not to be understood as an universal behavior: a wider study comprising over 100 research centres in France found a much lower per person research footprint of about 5 tons of equivalent CO$_2$, with different contributions such as purchases, commute, travel and heating having each comparable weight; this would be roughly comparable to their average footprint as consumers. It seems obvious that different research environments seem to favour different sources as the main contributions and, crucially, also very different total carbon footprints.\cite{EstevezTorres2024}

The Twelve Principles of green chemistry constitute a foundational guideline for the development of green chemistry. Principle 1 is Prevention, which emphasizes that it is preferable to prevent waste generation rather than dealing with its treatment or cleanup after it has been created. Principle 6 is Design for Energy Efficiency, and refers to energy requirements of chemical processes and how to minimize their environmental and economic impacts. \cite{Anastas2010}

In this context, there is little doubt that computer-based predictive tools are contributing to reduce CO$_2$ footprint. Currently, researchers can design safer chemicals and processes, anticipate potential environmental impacts, and optimize reactions for minimal energy and resources consumption while reducing waste generation. For instance, in the field of toxicology, by predicting their properties before the actual synthesis; thereby reducing the need for extensive experimental trials. \cite{Anastas2010, Mammino2023}. However, in the context of green chemistry, the environmental impact of resources and time-consuming computational modelling as opposed to low-tier computational models remains barely explored.\cite{EstevezTorres2024}

More recently, we have seen increased concerns regarding the carbon footprint from computations, \cite{Allen2022} and only lately have tools and guidelines been widely available to computational scientists to allow them to estimate their carbon footprint and be more environmentally sustainable \cite{GREENER2023,greentool}$^,$
During the peer review of this work, an excellent Tutorial Review appeared on this topic.\cite{Schilter2024}.
The exploding footprint of the so-called "generative artificial intelligence" is of course a cause for major concern from this perspective, and we should note that it also affects academia. Early studies are estimating a substantial and rapidly increasing fraction of academic content that is being produced by means of chatGPT.\cite{liang2024mapping}

Zooming momentarily out from scientific computing to computing in general, computing's global share of carbon emissions has been estimated to be from as low as 1.8\% if one focuses on operating costs to as high as 3.9\% if the full supply chain is taken into account, meaning it's comparable to air travel. \cite{freitag2021real} More importantly, while most economic sectors overall are starting to design or implement plans to reduce carbon emissions, computing's emissions are still strongly on the rise. This is despite continuous improvements in computational efficiency, including efforts towards green computing, as these have been consistently overtaken by increases in demand.\cite{knowles2022our} Indeed, emissions from computing, accounting for the production of the devices, have been projected to be close to 80\% of our emissions budget by 2040 to limit warming to 1.5$^\circ$C. \cite{vanderbauwhede2023frugal} This impossibility of continuing with ``business as usual'' has resulted in a call for ``frugal computing'', where computing is urgently recognized as a limited resource in the sense that it pushes against the hard boundaries of a hospitable planet. Ultimately -- but sadly this means sooner rather than later -- we need to advance towards zero-carbon computing, whether this means ``doing more with less'', or even ``doing less with much less''.

\section*{Frugal computing and frugal modelling}

These necessary but extremely ample goals need to be translated into each particular context. We will need to consider the shared costs, in terms of climate consequences, of the choices we make when we are doing science. We will focus herein on the open and challenging question of  translating frugal computing into frugal modelling. As a side remark, note that while the use of "generative AI" in academia risks bearing a higher carbon footprint should the current trend continue, it can be considered a solved problem, in the sense that the academic world has already demonstrated the ability to produce scientific literature without resorting to Large Language Models. In the matter of producing written text, the sane decision would be to continue with ``business as usual''. Thus, let us focus on what we refer to as frugal modelling, where we implicitly include any effort employing machine learning.

As we know, frugal computing focuses on minimizing hardware and energy resources while maintaining accessibility and affordability. Frugal modeling, in a similar vein, aims to develop models that require minimal computational power, data, and complexity. These models are designed to be efficient, simplified, and resource-conscious, allowing them to perform essential functions without excessive overhead or reliance on advanced infrastructure. Following the last ideas, we define frugal modeling as an approach that begins by considering the scientific question to be answered and the carbon footprint that is justifiable to emit in the process. When including the carbon footprint in the cost-benefit analysis, the goal is not only to optimize the efficiency of the method but also to limit the overall damage to the climate.\cite{Lannelongue2021} In other words, frugal modelling consciously avoids the "rebound effect": improvements in efficiency should not be offset by running significantly more calculations, which would result in an increasing overall carbon footprint. This was stated as "Rule 9: Be aware of unanticipated consequences of improved software efficiency" in Lannelongue et al's "Ten simple rules to make your computing more environmentally sustainable".\cite{Lannelongue2021} It is equally important to avoid perverse economic incentives, where sunk and fixed costs, such as the purchase of a supercomputer or the salary of a researcher, make it seem economically optimal to maximize the comparatively lower cost of keeping the supercomputer running at full capacity. Paradoxically, what can seem avoiding computational waste from the point of view of the system's administrator can actually be wasteful if the extra calculations performed to keep the supercomputer from being idle do not bring in any improved scientific insights.

In terms of implementation, frugal modeling emphasizes critical scientific thinking to minimize waste while pursuing knowledge of the highest available epistemic quality, if possible beyond a "justified true belief".\cite{Gettier1963} We do not need to settle for lower quality science, but we need to be held accountable for our carbon footprint, and this includes avoiding emissions that do now significantly improve the quality of the generated knowledge. Similar to the principles of green chemistry, where the use of problematic solvents in a method must be justified by demonstrating that no other viable alternatives exist for achieving the desired outcome, we propose that for models requiring thousands or tens of thousands of processor hours, a good faith analysis should be conducted to show that addressing the same problem with less harmful methods is unfeasible. Furthermore, a convincing justification must be provided that answering the specific scientific question in question warrants the associated climate impact. For complicated questions, it is often the case that in a sequence of stages taken to answer a scientific question, the one that we can improve by throwing more computing power at it is not the same one that limits the actual knowledge we can obtain. We will see an example below in the case study of spin states vs molecular distortions.

 Unfortunately, as a community, we have become accustomed to employ increasingly unsustainable amounts of computing resources in our calculations, meaning business as usual is not an option compatible with our societal commitments to a lesser climate catastrophe. Let us emphasize again here the fact that more efficient does not equal more frugal. There is a continuous striving for efficiency in computing, also in theoretical chemistry, see e.g.\cite{Steinbach2024} but as long as this is oriented to optimizing the return on investment, it is likely to produce increased emissions, as it has been happening historically, in this as in other economical sectors.\cite{Alcott2005} Fortunately, it is possible to answer plenty of interesting questions in Nanoscience, chemistry, Physics and Materials modelling while employing a frugal approach. In particular, we will focus here on our own field of expertise, namely magnetic molecules, although the general ideas may be extrapolated to many fields as well as to experimental chemistry.

Magnetic molecules have been studied for decades, firstly as controllable models for interactions and phenomena in solid-state Physics and more recently molecular nanomagnets have been presented as candidates for bits, \cite{Guo2018,goodwin2017molecular,Duan2022,Magott2022}
qubits, \cite{godfrin2017operating, gaita2019molecular}, 
p-bits, \cite{gutierrez2023lanthanide}
and also as components for nanotechnological devices.\cite{Mcadams2017,Long2020,Serrano2022,Aiello2022}
Manipulation of individuals spins, once a distant dream, is today a practical reality, if not one of immediate practical applicability. In parallel, advances have been made in modelling the influence of the chemical environment on said spin states and their dynamics, \cite{Rinehart2011,Takahashi2011,Ungur2016,Lunghi2017CS,lunghi2019phonons,Rosaleny2019,gu2020origins}
with wildly different computational costs, as we will see below in some detail. Indeed, a frugal approach is possible in this field thanks to the efforts over many years resulting in the development of analytical approaches and semi-empirical methods.\cite{staab2022analytic, malrieu2014magnetic}

Herein we present a couple of case studies focusing firstly on the contents of the models and secondly on their implementation. In the next section ``Choosing and solving affordable models'' we will present different alternatives for the modelling of the electric field modulation of the ligand field for the coherent control of the spin states in a molecular spin qubits.\cite{liu2021quantum} In the section ``Coding and running inexpensive implementations'' we will present a frugal model for magnetic molecules as probabilistic bits, \cite{gutierrez2023lanthanide} which can also serve, less frugally, to model their macroscopic magnetic properties.\cite{hu2023high} As we will see in these examples, models that can be similarly useful in practice can have costs varying in many orders of magnitude. Additionally, re-implementing an already frugal model to a more efficient implementation can significantly further the savings. Note however that unless frugality is maintained as a boundary condition, mere computational efficiency will often just lead to a rebound effect, i.e. increased use (precisely because of the improved return on investment) and increased emissions, same as in other sectors.\cite{Alcott2005} We will include some further context in the conclusions to clarify how the efforts we propose fit within the scope of green chemistry.

\section*{Choosing and solving affordable models}

The first questions that define a scientific work are typically ``what is the problem or question we aim to answer?'' and, closely related, ``how are we modelling this?''. An informal cost-benefit analysis follows, i.e. how many resources one needs to invest vs what one gets from it. In current academia, the resources or cost are related with the funding and computational capabilities of the research group, rather than an emissions budget. The benefits, when passed through the academic filter, are still about knowledge and the common good, but projected onto high-profile publications. There is a risk of cognitive biases favouring methods with a higher computational expense for no actual improvement in the generation of knowledge, as we exemplify below, and this leads to a waste we collectively cannot afford.

To illustrate this problem, let us focus on the different pathways that one can choose to model the effect of the ligand field on the magnetic and spectroscopic properties of metal ions, a question that has received some attention in the past decade in the context of the so-called single ion magnets and molecular spin qubits, since the spin dynamics of magnetic molecules are in large part based in the variation of the energies of the different spin states with distortions of the molecular structure.\cite{Guo2018,goodwin2017molecular,lunghi2017role,lunghi2019phonons,Rosaleny2019,gu2020origins,dergachev2023analytical,ullah2019silico}

The widely accepted standard in this field are \emph{ab initio} calculations, where complete active space perturbation theory (CASPT2) or n-electron valence state perturbation theory (NEVPT) are considered superior to the complete active space self-consistent field (CASSCF) for fundamental reasons, and MOLCAS or ORCA are employed as standard computational codes. A comparatively fringe modelling approach is based on effective charges acting on the f orbitals; this is widely considered much less exact, again for fundamental reasons. It is not often that the predictive power of the two tools is compared with the measuring stick of experimental spectroscopic information, but at least in one example where this was done, we found no clear benefit in the extra computational cost of using more sophisticated models since, for the task of predictively estimating energy-level distribution, including the energy of the first excited state, CASPT2 did not prove to be superior to CASSCF and CASSCF was not found to be superior to the radial effective charge (REC).\cite{baldovi2016rational} Wider and very critical reviews have also found a similar trend, when comparing effective theories vs ligand-field theory vs ab initio calculations, in the sense that neither of the approaches is a good fit for experimental results. This means both kinds of methods demonstrably fail at allowing us a high quality knowledge, although they do so in different ways. Effective theories can miss important parts of the Physics and high-level ab initio calculations tempt us to lose a critical perspective.\cite{van2015comprehensive} It has been argued that CASxxx methods in particular have to be considered qualitative with respect to magnetochemical properties.\cite{van2015comprehensive}
This is indeed a general problem when striving for a frugal approach: the need for benchmarking, which ideally should be done with experiments rather than with another theoretical method.

\subsection*{Case study: Spin states vs molecular distortions}

The coupling between spin states and vibrational excitations, generally detrimental due to its role in the decoherence, also allows exploiting spin-electric couplings for quantum coherent control of a qubit. To decipher the origin of the decoherence mechanism, it is necessary to determine the spin-vibrational couplings or vibronic couplings for each vibrational mode. As vibrational coordinates are orthonormal, they provide a good basis set to evaluate spin-electric couplings (SECs).

Herein, we present a computationally inexpensive methodology to explore both vibronic couplings and SECs. This computational methodology consists of three steps, the first step is to determine the spin energy levels at multireference level (e.g., CASSCF with spin-orbit coupling (CASSCF-SO)) in crystal geometry, the obtained energy levels are employed to parameterize the ligand field Hamiltonian within REC model implemented in the SIMPRE code \cite{baldovi2012modeling, baldovi2013simpre}. Alternately, experimental spectroscopic information can be used for this step. In the second step, the geometry is optimized using density functional theory (DFT) to determine the vibrational frequencies and their corresponding displacement vectors. 
To determine vibronic couplings, the final step consists of generating distorted geometries along the normal vectors and employing the REC model to determine the spin energy levels and crystal field parameters (CFPs). To estimate SECs, an additional step is required where dipole moment is determined along a vibrational coordinate at DFT level to construct a new charge affected vibrational basis; alternately, this can be obtained inexpensively by employing effective charges. In a final step, spin levels and CFPs are determined using the REC model.

This process is computationally very demanding when performed solely at \emph{ab-initio level} instead of with an effective charge model. We applied this scheme to spin-qubit candidate \HoWF (in short \HoW) and compared the vibronic couplings  and SECs with those already determined by solely using ab initio level\cite{liu2021quantum, blockmon2021spectroscopic}. The equilibrium spin energy levels and wave function composition of \HoW are provided in Supplementary Table 1.
This scheme has already proven effective for determining key vibrations responsible for spin relaxation in molecular nano-magnets\cite{ullah2019silico}.

\begin{figure}
\begin{center}
\includegraphics[width=8.0cm]{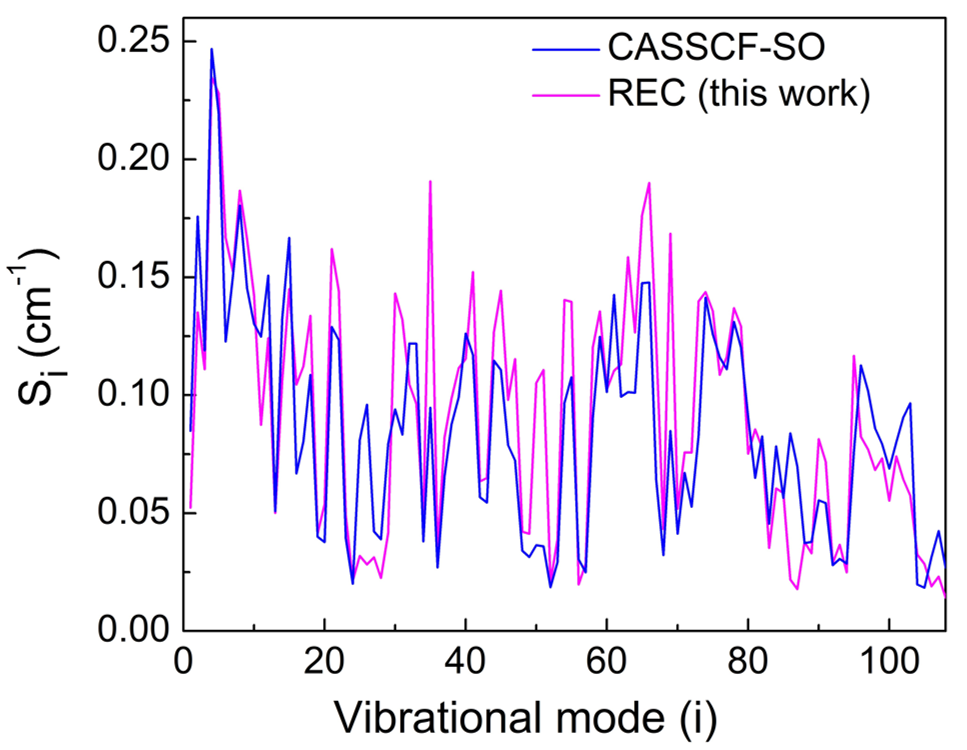}
\caption{Vibronic coupling strength
S$_i$, calculated for each vibrational frequency of \HoW using both CASSCF-SO method and REC model  \cite{blockmon2021spectroscopic}. Note that the modes are merely ordered by increasing energy and have not been reordered by similarity of the displacement vectors.}
\label{SPC}
\end{center}
\end{figure}

The vibronic couplings are obtained by distorting the equilibrium geometry along each normal mode coordinate ($x_i$). The evolution of each crystal field parameters (CFPs) was fitted into a second-order polynomial, the first derivative versus $x_i$, allowing us to determine vibronic couplings for each normal mode, i.e. $\left(\frac{\partial B_k^q}{\partial x_i}\right)_0$. The overall effect can be obtained by averaging over different ranks ($k$, $q$) of CFPs, as in eq. \ref{eq:vib} \cite{chang1982optical}.
\begin{equation}\label{eq:vib}
	S_i = \sqrt{\frac{1}{3}\sum_{k}\frac{1}{2k+1}\sum_{q=-k}^{k} |\left(\frac{\partial B_{k}^q}{\partial x_i}\right)_0}|^2
\end{equation}
The obtained vibronic couplings of each vibrational mode are plotted in Fig. \ref{SPC} and compared with previously determined using CASSCF-SO method. The vibronic couplings obtained using the REC model differ, on average, by $\pm$0.022 cm$^{-1}$ from CASSCF-SO results, an overall satisfactory agreement where nevertheless one finds substantial relative deviations in several vibrational modes. Differences between the estimates of the two models are generally attributed to not considering the second coordination sphere of \HoW in the REC model. Nevertheless, the overall comparison is satisfactory. The detailed values of vibronic for each vibrational mode is provided in Supplementary Table 2.

To seek insight into spin-electric couplings, we established a relation between spin Hamiltonian and molecular distortion as a function of the dipole moment. The spin-electric couplings are defined as shifts in transition frequency ($\delta f$) between two spin energy levels. The relationship between applied electric field (\Efield) and spin state is determined by noting that \Efield will cause a change in the dipole-moment ($\delta p$) of the molecule, lowering the electric potential. The stabilization via the electric potential is exactly compensated by the elastic cost of distorting the molecular structure, i.e. $E\cdot\sum_i\delta p= \sum_i\kappa_i\delta x_i$. Thus, by calculating the electric dipole moment as a function of the mode displacements, we can quantitatively extract the displacements as a function of the applied \Efield. Each normal mode is associated with force constant $\kappa_i$ and reduced mass $\mu_i$ (yielding eigenfrequency $\omega_i = \sqrt(\kappa_i/\mu_i)$. The electric dipole $p$ depends on the displacement of modes $x_i$ , and this determines the coupling of the mode to an applied \Efield or to incident light, that is, its infrared intensity. By linear combination of all normal modes, we can find an effective displacement as a function \Efield, e.g., $x_{eff}=\sum_i^{3N-6} x_{i}(E)$. 

\begin{figure}
\begin{center}
\includegraphics[width=8.6cm]{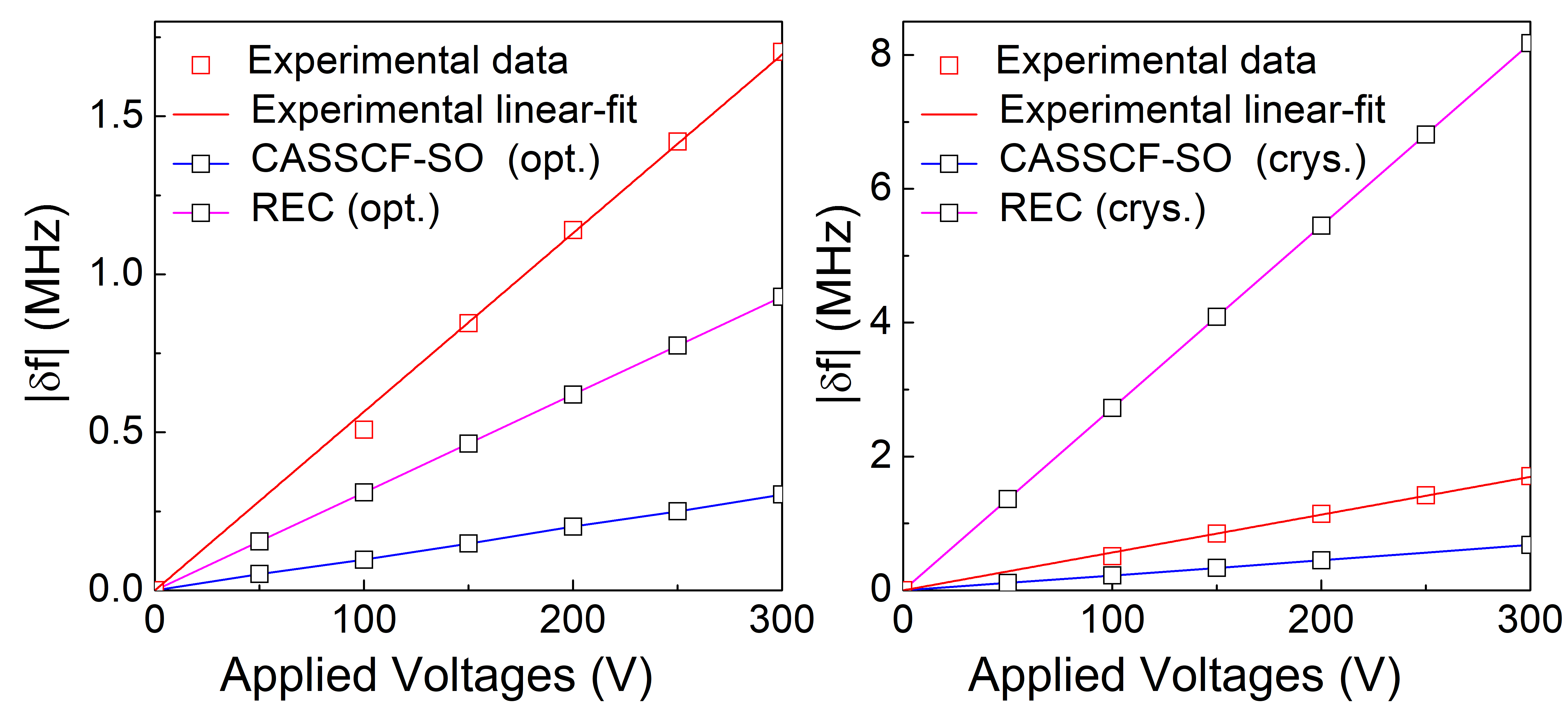}
\caption{The shift in transition frequency (\df) versus applied  voltage V, showing a linear \Efield coupling in \HoW. Calculations correspond to the DFT-optimized structure (left) and to the crystallographic structure (right). \cite{liu2021quantum}}
\label{sec}
\end{center}
\end{figure}

Note that this model, while reasonable, relies on a series of approximations. The molecular structure, electric dipole and vibrational modes obtained by DFT do not correspond exactly to what happens within the crystal, e.g. we are not considering the displacement of crystallization water molecules and counterions and neither are we considering the distortion of the orbitals as a result of the electric field. Within this framework, one can estimate how the spin state has evolved as a function of, $x_{eff}$ using the REC model described above to determine the $\delta f$. The results are shown in Fig. \ref{sec} and compared with SECs determined at CASSCF-SO. We repeated  this process for both DFT optimized geometry and crystallographic geometry for different applied \Efield (the distance between two plates where the sample is placed one could convert \Efield to Voltage units (V)).  

The linear increment in transition frequency was observed for both optimized and crystal geometries, which is in accordance with experiment and previously determined CASSCF-SO level. 
From Fig. \ref{sec}, one can see that the REC values are very sensitive to the geometry, whereas CASSCF are comparatively stable. 
But the overall tendency is well reproduced and satisfactory and moreover both methodologies are similarly inaccurate: for the optimized structure both techniques underestimate the shift, with REC being off by a factor $<2$ and CASSCF-SO being off by a factor of $>5$, whereas for the crystal structure they fail in different directions, and it is REC the one which is off by a factor of $<5$ and CASSCF is off just by a factor of $>2$. In any case, with this linear progression spin-electric couplings constant in units of Hz/Vm$^{-1}$ are provided in Table \ref{tab:sec}, both methodologies resulted in same order of magnitude for SECs constant. The reason for the qualitative coincidence between the two very different models of the $\delta$f vs voltage is likely the fact that neither is perfect but they are both good enough, and their exactness is actually limited by the many approximations in the previous parts of the model as detailed above. The detailed values of SECs for different voltages are provided in Supplementary Table 3.
\begin{table*}
	\begin{center}
	\caption{Spin-electric coupling constant (SEC) of \HoW are computed using both  CASSCF-SO method and REC model for crystalline and optimized geometry and compared with experimentally determined constant \cite{liu2021quantum}.}\label{tab:sec}
		\begin{tabular}{lccccc}	
\toprule		
   & Exp.  & CASSCF-SO (opt.) & CASSCF-SO (crys.) & REC (opt.)  & REC (crys.)\\
\midrule
SEC (Hz/Vm$^{-1}$)  & 11.4  & 2.0     & 4.5   & 6.2   & 54.5 \\
\bottomrule	
	\end{tabular}
	\end{center}
\end{table*}

For optimization and vibrational frequency calculations, the computing time was $\approx$ 160 hrs using Density function theory (DFT) implemented in Gaussian16. The spin-energy levels, using ab-initio approach (either CASSCF of the restricted active space state interaction (RASSI)) method implemented in Molcas, the processing time of $\approx$ 10.33 hrs was spent using 4 i9 processors in MPI parallel processing and 64 GB of memory on a local server. 
The determination of spin energy levels for spin-electric coupling at fully ab-initio approach, a total of 13 geometries including equilibrium was calculated, an approx. of 134.33 hours processing time was used. For vibronic couplings, 135$\times$6$\times$10.333 (no. of modes $\times$ no. of geoms. $\times$ processing time) $\approx$ 8370 hrs of computation.
When the corresponding task was performed using the REC model implemented in SIMPRE, the total processing time to determine spin-electric couplings, was 13x1=13 seconds and for spin-phonon couplings, 135x6x1=810 seconds (13.5 mints).

The associated energy expense of the \emph{ab initio} approach (considering the electricity cost, rather than the full supply chain) is about 1 MWh, with a carbon footprint which can be estimated to be (assuming the average energy mix in Spain) in the order of 200 kg of CO$_2$ equivalents, see Table \ref{tab:comTime} for detailed computational time and energy cost at each step.\cite{lannelongue2021green} This is comparable to a thousand  
km in a passenger car, or similar to the per-passenger emissions of a medium-distance flight.
This is not an absurd cost, but it is not environmentally negligible, either. When intensive calculations result in carbon footprint comparable to those of flying, this cost should be taken into account by environmentally-conscious researchers.~\cite{Allen2022} In the REC approach, in contrast, the SEC estimation would have a negligible carbon footprint about $10^4$ times smaller. Actually, in that case the total cost would be dominated by the initial DFT cost of structure optimization and calculation of the vibrational modes, so the actual factor in the savings is about 50 i.e. less than 5 kg CO$_2$ equivalents, enough to consider the inexpensive method as environmentally acceptable. 

\begin{table*}[htbp]
  \centering
  \caption{Computational time in hours (hrs) and corresponding energy consumption in kilo-watt-hour (kWh) for different computational task}\label{tab:comTime}
    \begin{tabular}{p{12.715em}rr}
    \toprule
    \multicolumn{1}{r}{} & \multicolumn{1}{l}{
    Computation time (hrs)} & \multicolumn{1}{l}{{Energy consumed (kWh)}} \\
    \midrule
    \multicolumn{1}{l}{DFT (opt, freq)} & 160   & 26 \\
    \multicolumn{1}{l}{CASSCF-SO} & 10 & 1.27 \\
    \multicolumn{1}{l}{REC} & 0.00028 & 0.00005 \\
    \midrule
    \multicolumn{1}{l}{\textit{\textbf{Ab-initio/Molcas}}} &       &  \\
    Spin-electric couplings & 294 & 43 \\
    Spin-vibrational couplings & 8540 & 1056 \\
    \midrule
    \multicolumn{1}{l}{\textit{\textbf{REC/SIMPRE}}} &       &  \\
    Spin-electric couplings & 170 & 28 \\
   Spin-vibrational couplings & 171 & 28 \\
    \bottomrule
    \end{tabular}
\end{table*}

Even more promising, an analytical scheme was recently demonstrated to estimate nonadiabatic coupling and state-specific energy gradient for the crystal field Hamiltonian describing lanthanide single-ion magnets. \cite{dergachev2023analytical} Within this scheme, a single-point calculation of the desired accuracy -or experimental spectroscopic information- can be employed to fine-tune parameters of a very inexpensive model, which then allows taking analytical derivatives for any desired perturbation in the molecular geometry, therefore saving hundreds of calculations that would be required to estimate the same derivatives numerically. As a result, even within frugal modelling constrictions, it is possible to model spin relaxation using sophisticated nonadiabatic molecular dynamics. This clearly points towards the right direction we need to follow to keep producing good science that is compatible with the planetary boundaries: we need to know our systems well, find or develop a minimal model that recovers a good part of the relevant physics, and then, if possible, employ an analytical approach to solve the problem, rather than brute-force throwing computational power into it.

\section*{Coding and running inexpensive implementations}

An additional question, beyond the model we choose to solve, is how we actually solve it. The ease of use of programming languages has been being increasing over time, with a rising popularity of high-level computing environments. These provide interactive exploratory environments that make language features and libraries immediately available to scientists who can use them to explore a problem domain. This contrasts with the classical edit/compile/run cycle of C or Fortran programming, which typically requires separate computation and post-processing/visualization steps. \cite{Pérez2011} Additionally, the popularity of high-level languages means solutions to syntax errors and general advice can be found on the Internet. High-level programming often comes at a cost in resource, but at the same time, these languages are designed to be easy for humans to read and write, allowing developers to focus more on problem-solving rather than worrying about low-level details like memory management or hardware specifics. Nevertheless, they can support green coding practices by facilitating efficient coding techniques and resource management. While high-level languages are often not as efficient as low-level ones (e.g., C or assembly), they encourage code practices that can still lead to efficient performance through optimized libraries, efficient algorithms and data structures, contributing to lower energy consumption.

\subsection*{Case study: different implementations of STOSS}

Let us take an established physical model and deal only with the implementation choice. 
We now explore an already published program that models molecular nano-magnets and their application as probabilistic bits, called STOSS, \cite{gutierrez2023lanthanide} and test a different implementation of the same algorithm. 

\begin{table*}[htb!]
  \centering
  \caption{Time required to process the evolution of a lanthanide-based, molecular 2 p-bit network where each p-bit is embodied by 1 million spins and the simulation runs for 10 thousands time steps. \cite{gutierrez2023lanthanide}}
    \begin{tabular}{p{25em}ccc}
    \toprule
    \multirow{2}{*}{Operation} & \multirow{2}{*}{File} & \multicolumn{2}{c}{Time in sec} \\
     & & Matlab & Python \\
    \midrule
    Read input data from EXCEL file & user configurations & 0.2346 & 0.2080 \\
    \midrule
    Read system characteristics from EXCEL file & read\_data & 0.0725 & 0.0094 \\
    \midrule
    Calculate magnetic relaxation &mag\_relaxation & 0.0021 & 0.0010 \\
    \midrule
    Calculate probabilities of each spin to flip in the 1st p-bit & Bolztmann\_distribution & 0.0010 & 0.0010 \\
    \midrule
    Iteration process ("for" loop) for the 1st p-bit & changeable\_field & \textbf{305}   & \textbf{16800} \\
    \midrule
    Calculate probabilities of each spin to flip in the 2nd p-bit & Bolztmann\_distribution & 0.0473 & 0.2630 \\
    \midrule
    Iteration process ("for" loop) for the 2nd p-bit & changeable\_field & \textbf{275}   & \textbf{16900} \\
    \midrule
    Average p-bits states over time & mean\_matrix\_state & 0.0460 & 0.0036 \\
    \midrule
    Association analysis between both p-bits & association & 0.1484 & 2.3946 \\
    \midrule
    Plotting results & plotting & 0.7876 & 0.4762 \\
    \midrule
     Total time & & \textbf{581}   & \textbf{33700} \\
    \bottomrule
    \end{tabular}%
  \label{tab:t2}%
\end{table*}

STOSS consists in a custom implementation of a Markov Chain Monte Carlo algorithm for each of the $N$ independent particles (in this case, effective spins $S=1/2$). The relative Markov chain probabilities for the spin flip between ground and excited spin states correspond to the relative Boltzmann populations of the two effective spin states $M_S=+1/2$, $M_S=-1/2$. Each computational step has an associated natural time duration that is derived from parameterized average spin dynamics, thus the model allows one to follow $N$ independent time trajectories. There are three main scenarios studied using experimental data for comparison, and all the details could be found in the Supporting Information File of Gutiérrez-Finol et al \cite{gutierrez2023lanthanide}.

Each implementation of the same program, even if following a given algorithm as closely as possible, have distinct costs, in terms of memory use, runtime and energy consumption. This has often been analyzed in particular for implementations employing different programnming languages.~\cite{Pereira2017}
In the case of Matlab vs Python a major difference arises in "for" loops, where Matlab is faster than Python. More generally, M language has a strongly typed syntax, often resulting in a improvements in memory usage and processing time. Identifying the type of each variable at compile-time allows the compiler to optimize the code, saving time and being able to use the minimum amount of memory. 
That being said, we are not claiming here that the behaviour we present here is univocally associated with implementing the model in python vs in M language, since many different approaches are always possible even within a given language and algorithm.

For this research the study was carried out using a desktop computer (Processor 11th Gen Intel(R) Core(TM) i9-11900K @ 3.50GHz, and installed memory of 16.GB with 15.8GB available) with Windows 10 Enterprise (22H2 version). The following Python (3.10 64-bit)  modules were used: NumPy/Scipy (using Intel Math Kernel Library extension), Matplotlib, Pandas, Collections, random, math, and time. On the other hand, we used Matlab R2023b, 64-bit. 

We performed the calculations on the third case of study in a lanthanide-based, molecular spin p-bit network \cite{gutierrez2023lanthanide} which corresponds to the longest computation in the paper. Having in mind the original case, we implemented the simulation maintaining the same conditions in both scenarios, therefore, we explored a 2-p-bit architecture where each p-bit is constituted by the collective signal of $10^{6}$ magnetic molecules which evolve freely, with the molecules corresponding to the first p-bit evolve in absence of a magnetic field and the one corresponding to the second p-bit evolve in presence of a magnetic field determined by the state of the first p-bit. The program is divided in seven functions where each one accomplishes a specific task, two functions are reused for the calculation of the probability of each p-bit to change its state. Table ~\ref{tab:t2} presents the time speed processing for each function and as it could suggest by far the largest difference is from the "for" loops when the program iterates over each spin at each time step. 

In this case the overall total runtime is rather short in any case, as are the associated emissions. However, should one employ STOSS to fit experimental data, as was done recently,~\cite{hu2023high} one would be tempted to explore a wide parameter range, and that would give rise to a larger carbon footprint if the more expensive implementation was used. A good approach here would be, as soon as the expected calculation time veers into the hundreds or thousands of hours, start thinking about a less expensive implementation. Of course it can be even better to optimize the model rather than just the implementation, and we are indeed working on that. Additionally, we now present an updated version of the program, implementing several optimizations to enhance performance. By leveraging the vectorized operations and built-in functions available in the MATLAB programming language, we significantly reduced computational time. These optimizations allow for more efficient data handling and processing, minimizing the need for iterative loops. As a result, the program is now capable of delivering faster results without compromising accuracy, offering an overall improvement in both speed and functionality.

%
%

\section*{Conclusion and context}
\label{conclusion}


The most general question we are dealing with here is how to take into account the goal of environmental sustainability when doing science; in the case of chemistry this is the field of green chemistry. A closely related question is how to balance environmental sustainability and scientific advancement. Our proposal is to aim for a rational cost-benefit analysis, and a prerrequisite for this is being aware of the costs. Only after researchers, research groups and research institutions are quantitatively aware of how their carbon footprints derives from their choices, we will be able to rationally reduce emissions.

Quantifying CO$_2$ footprint in chemistry laboratories is far from a trivial task and requires and exhaustive life cycle assessment (LCA). We should account for everything when assessing an LCA: materials acquisition and input, manufacturing and production, packaging and distribution, product use, disposal and recycling).\cite{Anastas2000} Consequently, in the framework of green chemistry, we urge to implement objectives of sustainability also into computational chemistry research, not only to design safer chemicals and processes to avoid physical experiments, but also to reduce energy consumption and minimize waste. Just as we aim to prevent and minimize when designing a chemical reaction, we should also standardize preventive practices during computational experiments.

Here we need to be aware of the urgency of the climate crisis and thus avoid relying on unproven solutions and risky bets on techno-optimistic futures. In particular, in the field of theoretical and quantum chemistry, it would be irresponsible to wait until some supposedly energy efficient quantum computers start solving practical problems. Presently, both the operation of quantum computers and the simulation of quantum circuits carry a considerable carbon footprint and are not being used to avoid the footprint of conventional supercomputers.\cite{Chen2023,Li2023} The same can be said of AI-based solutions, which may well bring increased productivity but have so far substantially increased the overall computing carbon footprint, mainly through the training cost of the models, with no clear path to net reduction.

We presented here two particular examples from the field of computational materials science illustrating an extremely common situation: when confronted to a calculation with a large carbon footprint, one can often choose either to solve a different model or to solve the same model via a different implementation, and obtain significant savings in carbon footprint without making any significant sacrifices in scientific yield. Currently this is mostly overlooked, and if anything the most expensive methods tend to enjoy a higher prestige and are considered more trustworthy. Often, this means that wasteful methods allow for easier or better publishing venues.Indeed, this is a general problem rather than exclusive for computation: a wasteful excess of experimental techniques is rarely if ever seen as a problem from the publishing perspective, and only a matter of money.

In our case, the research presented in Liu et al \cite{liu2021quantum} was originally submitted with the frugal method, but during the refereeing process we were requested to switch to the method that is in principle more exact but which we show here is actually wasteful in this case. 
As a community we need to do better, and the main factor is choosing affordable models to solve problems, with a minor but sizeable contribution of finding an inexpensive way to implement these models. Crucially, we need to beware of the Jevons' paradox or rebound effect,~\cite{Alcott2005} meaning a more efficient method, if not coupled to resource consciousness, by itself leads to an increased usage which often overshoots the savings. This can be seen as a particular case of the necessary attitudinal change in all of chemistry, ~\cite{Kummerer2020}. Thus, we call herein for "frugality" rather than for "efficiency".
More generally, just as we consider the ethical repercussions of animal experimentation, or dealing with patient data, eventually we will need to include the risk of carbon footprint wastefulness as an ethical concern in research.

\section*{DATA AVAILABILITY}
The data reported in the first part of this work are available as part of the Supplementary Information.
For the second part, all custom data generated and employed for this study are available at  \href{https://github.com/gerlizg/STOSS-MATLAB}{https://github.com/gerlizg/STOSS-MATLAB}.

\section*{CODE AVAILABILITY}

For the second section of this paper, the original code (python version) named STOSS (for STOchastic Spin Simulator) is available at \href{https://github.com/gerlizg/STOSS}{https://github.com/gerlizg/STOSS}. The new version of the program using Matlab and the instructions to reproduce all the results could be found in the Supporting Information Section S2 and at \href{https://github.com/gerlizg/STOSS-MATLAB}{https://github.com/gerlizg/STOSS-MATLAB}. Finally, the optimized version of the simulator could be found at \href{https://github.com/gerlizg/STOSS-Optimized-Matlab-Version} {https://github.com/gerlizg/STOSS-Optimized-Matlab-Version}.

\bibliographystyle{apsrev4-1}
\bibliography{arxiv}

\section*{Acknowledgements}
A.G.A. has been supported 
by the European Union (EU) Programme Horizon 2020 (FATMOLS project), and by the Generalitat Valenciana (GVA) CIDEGENT/2021/018 grant. 
A.G.A. thanks grant PID2020-117177GB-I00 funded by MCIN/AEI/10.13039/501100011033 (co-financed by FEDER funds). 
This study is part of the Quantum Communication programme and was supported by grant PRTR-C17.I1 funded by MCIN/AEI/10.13039/501100011033 and European Union NextGenerationEU/PRTR, and by GVA (QMol COMCUANTICA/010).

\section*{Conflict of Interest}

The authors declare no competing interests.

\end{document}